\documentclass[pra,aps,twocolumn,showpacs,amsmath,amssymb,nofootinbib]{revtex4}
\usepackage{bm,graphicx,amsmath}
\usepackage{bbm}

\usepackage{bm}
\usepackage{amssymb}
\usepackage{epsfig,times}
\usepackage{epstopdf}
\usepackage{times}
\usepackage{txfonts}
\usepackage{amstext}
\usepackage{latexsym}
\usepackage{hyperref}

\begin{document}

\title{Entanglement detection in hybrid optomechanical systems}

\author{Gabriele De Chiara$^1$, Mauro Paternostro$^2$, and G. Massimo Palma$^3$}
\affiliation{$^1$ F\'isica Te\`orica: Informaci\'o i Fen\`omens Qu\`antics, Universitat Aut\`{o}noma de Barcelona, E-08193 Bellaterra, Spain\\
$^2$Centre for Theoretical Atomic, Molecular and Optical Physics, School of Mathematics and Physics, Queen's University, Belfast BT7 1NN, United Kingdom\\
$^3$NEST, Istituto Nanoscienze-CNR and Dipartimento di Fisica, Universit\`a degli Studi di Palermo, Via Archirafi 36, I-90123 Palermo, Italy}

\begin{abstract}
We study a device formed by a Bose-Einstein condensate (BEC) coupled to the field of a cavity with  a moving end-mirror and find a working point such that  the mirror-light entanglement is reproduced by the BEC-light quantum correlations. This provides an experimentally viable tool for inferring mirror-light entanglement with only a limited set of assumptions. We prove the existence of tripartite entanglement in the hybrid device, persisting up to temperatures of a few milli-Kelvin, and discuss a scheme to detect it.
\end{abstract}

\date{\today}

\maketitle

\section{Introduction}
\label{sec:intro}
Our ability to experimentally enforce quantum mechanical behaviors in complex systems of a diverse nature is improving at a steady pace~\cite{macro!}  and is swiftly leading us toward the grasping of the long craved quantumness at the macro scale. However, the faithful direct inference of quantum properties of parts of a complex systems or of a meso-system or macro-system is not a simple task to accomplish. This could be due to the intrinsic impossibility to directly address the system we are interested in or to its inherent complexity, which might face the lack of stringent and certifiable criteria for non classicality. The idea of  indirectly probing the system of interest by coupling it to fully controllable detection devices appears very appealing~\cite{Probes,MauroPRL}. It has spurred interest in designing techniques for the extraction of information from noisy or only partial data-sets gathered through observations of the detection system. The main problem in such an approach is embodied by the set of extra assumptions that one has to make on the mechanisms to test and the retrodictive nature of the claims that can be made. In fact, one can a posteriori interpret a data-set and infer the properties of a system that is difficult to address by post-processing the outcomes of a few measurements. However, this requires assuming knowledge of the working principles of the effect that we want to probe.

\begin{figure}[t]
\includegraphics[scale=1.3]{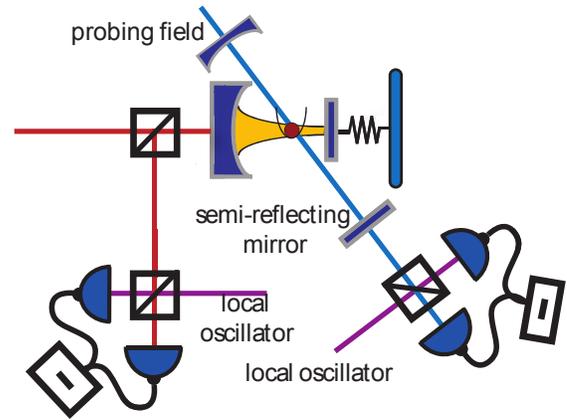}
\caption{
Sketch of the detection scheme. The probe field has the same periodicity of the cavity field. It impinges on a semi-reflecting mirror, creating a standing wave. Part of the probe light is transmitted and then measured with the output light of the cavity. 
}
\label{fig:sketch}
\end{figure}

In this sense, it would be highly desirable to conceive indirect detection strategies providing a faithful picture of the property to test by, for instance, ``copying'' it onto the detection stage, which would then be subjected to a direct estimate process. By focussing on a hybrid optomechanical device comprising an atomic, mechanical and optical mode, we propose a diagnostic tool for entanglement that operates along the lines of such desiderata.  We show that the entanglement established between the mechanical and optical modes by the means of radiation-pressure coupling can indeed be ``written" into the light-atom subsystem and directly read from it by means of standard, experimentally friendly homodyning. Our analysis goes further to reveal that such a possibility arises from the non-trivial entanglement sharing properties of the system at hand, which is indeed genuinely tripartite entangled. Very interestingly, we find that while any bipartite quantum correlation is bound to disappear at very low temperature, the multipartite content persists longer against the operating conditions, as a result of entanglement monogamy relations. Our analytical characterization considers all the relevant sources of detrimental effects in the system, is explicitly designed to be readily implemented in the lab and provides a first step into the assessment of multipartite entanglement in a macro-scale quantum system of enormous experimental interest. 

This paper is organized as follows. In section \ref{sec:model} we describe the physical system we consider and the equations of motion for the relevant variables. In section \ref{sec:ent} we show the results for the stationary entanglement shared by the cavity field, the BEC and the optomechanical oscillator. In section \ref{sec:det}, we discuss how the entanglement between the three different subsystems can be experimentally measured and, finally, in section \ref{sec:conclusion} we summarize and conclude.

\section{The physical model}
\label{sec:model}
The system we consider, sketched in Fig.~\ref{fig:sketch}, is made of a cavity where one of the two mirrors is free to move in the direction parallel to the cavity axis. In absence of radiation pressure from the cavity, the mirror moves harmonically with a frequency $\omega_m$ and is at equilibrium at a temperature $T$. The cavity is pumped through the fixed mirror by a laser of strength $\eta$ and is coupled to an intra-cavity BEC.
We assume the latter in the weakly interacting regime~\cite{StringariPitaevskii} where we can expand the field operator in a classical part (representing the condensate wave-function) and a quantum part representing quantum fluctuations. These are conveniently expressed in terms of Bogoliubov modes. It has been recently shown experimentally~\cite{Esslinger2008,Stamper2008} that the only Bogoliubov mode which is significantly coupled to the cavity is the one with momentum $2k_c$ where $k_c$ is the cavity mode momentum. We also assume ultralow temperature of the condensate so that thermal fluctuations are negligible and the BEC wave-function is not affected by the coupling to the cavity field. We write the Hamiltonian of the system made of the cavity field, the movable mirror and the BEC as
\begin{equation}
\hat{\cal H}=\hat{\cal H}_M+\hat{\cal H}_C+\hat{\cal H}_A+ \hat{\cal H}_{AC}+ \hat{\cal H}_{MC}
\end{equation}
where the mirror, cavity and BEC Hamiltonians are given respectively by
\begin{eqnarray}
&\hat{\cal H}_M&=\frac 12 m \omega_m^2 \hat q^2{+}\frac{\hat p^2}{2m},
\\
&\hat{\cal H}_A&=\hbar \Omega \hat c^\dagger \hat c,
\\
 &\hat{\cal H}_C&=\hbar (\omega_C-\omega_L)\hat  a^\dagger \hat a -i \hbar \eta (\hat a-\hat a^\dagger).
\end{eqnarray}
Here, $\hat q$ and $\hat p$ are the mirror displacement and momentum operators while $m$ and $\omega_m$ are its effective mass and oscillation frequency. The cavity (pump laser) frequency is $\omega_C$ ($\omega_L$)  and $\hat a$ is the cavity mode annihilation operator. The term proportional to $\eta{=}\sqrt{2\kappa{\cal R}/\hbar\omega_L}$ describes the pumping of the cavity and depends on the laser power ${\cal R}$ and the cavity decay rate $\kappa$. Finally, $\Omega$ and $\hat c$ are the frequency and bosonic annihilation operator of the Bogoliubov mode [cfr. Fig~\ref{fig:sketch}].
For small mirror displacements, the mirror-cavity interaction Hamiltonian can be written as~\cite{MauroNJP}:
\begin{equation}
\hat{\cal H}_{MC}{=}{-}\hbar \chi' \hat q \hat a^\dagger \hat a.
\end{equation}
It is useful to introduce the quadrature operators for the Bogoliubov mode
\begin{eqnarray}
\hat Q=\frac{\hat c{+}\hat c^\dagger}{\sqrt 2},
\quad
\hat P=i\frac{\hat c^\dagger{-}\hat c}{\sqrt 2},
\end{eqnarray}
so that the BEC-cavity interaction reads
\begin{equation}
\label{eq:HAC}
\hat{\cal H}_{AC}{=}\hbar \frac{g^2 N_0}{2\Delta_a} \hat a^\dagger \hat a + \hbar \zeta  \hat Q \hat a^\dagger \hat a,
\end{equation}
where $g$ is the vacuum Rabi frequency, $N_0$ is the condensate fraction, $\Delta_a$ is the detuning of the atomic transition from the cavity frequency and the coupling rate $\zeta$ depends on the Bogoliubov mode-function~\cite{PaternostroPRL2010}. The same model can be realized with two BECs trapped in the same cavity, where the mechanical oscillator is replaced by a second BEC. This setup can be realized straightforwardly as an extension of current experiments~\cite{Esslinger2008,privcomm}. It is also worth stressing that, although our model shares, at first sight, some similarities with the one considered in Ref.~\cite{genes}, it is distinctively different. First, the atomic media utilized in the two cases are different, with Ref.~\cite{genes} considering the bosonized version of the collective-spin operator of an atomic ensemble. Second, the resulting coupling Hamiltonians are built from quite distinct physical mechanisms. Third, the working conditions to be used in the two models in order to achieve interesting structures of shared quantum correlations are very different. We will comment on this latter point later.
\begin{figure*}[t]
{\bf (a)}\hskip6cm{\bf (b)}\hskip5cm{\bf (c)}
\\
\includegraphics[scale=0.46]{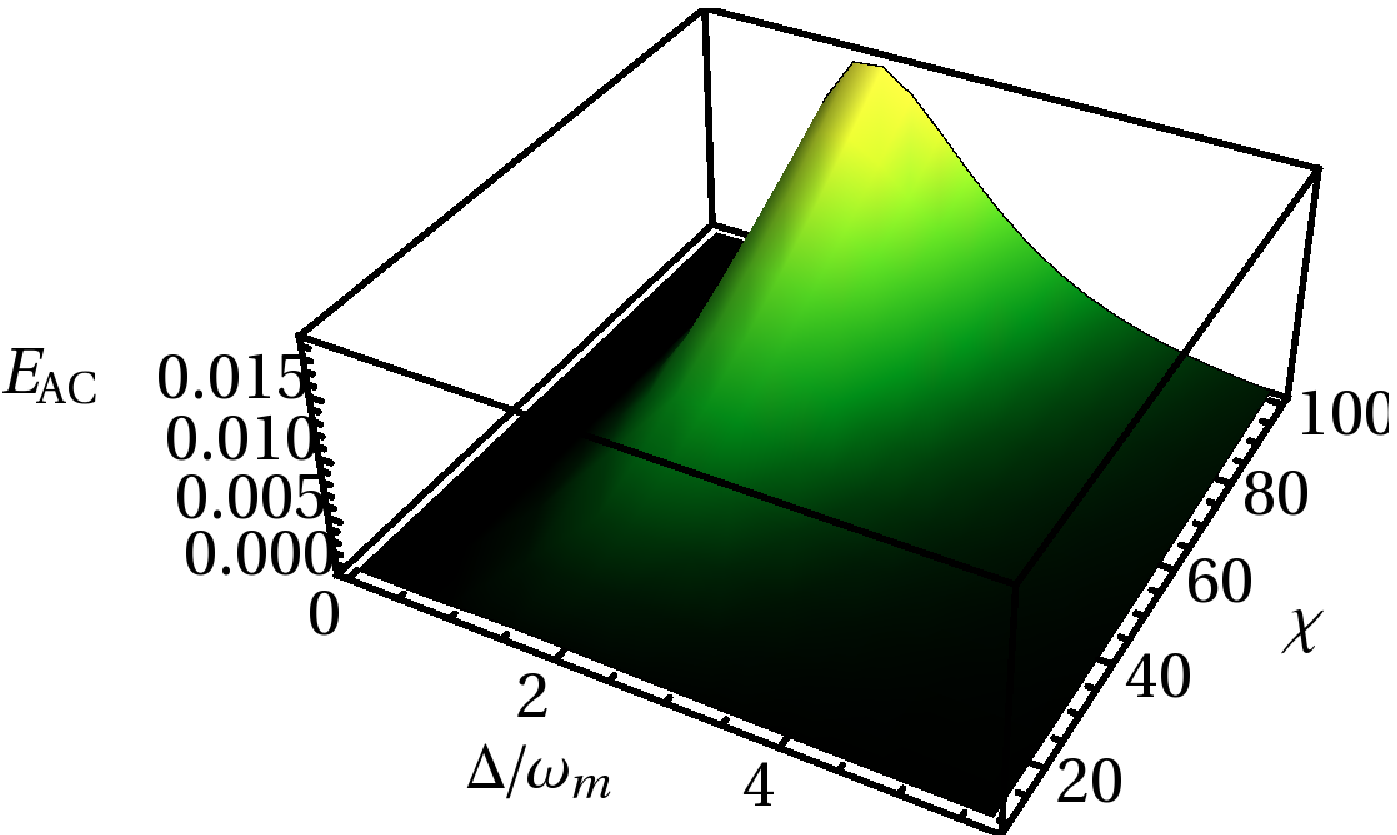}
%
\includegraphics[scale=0.52]{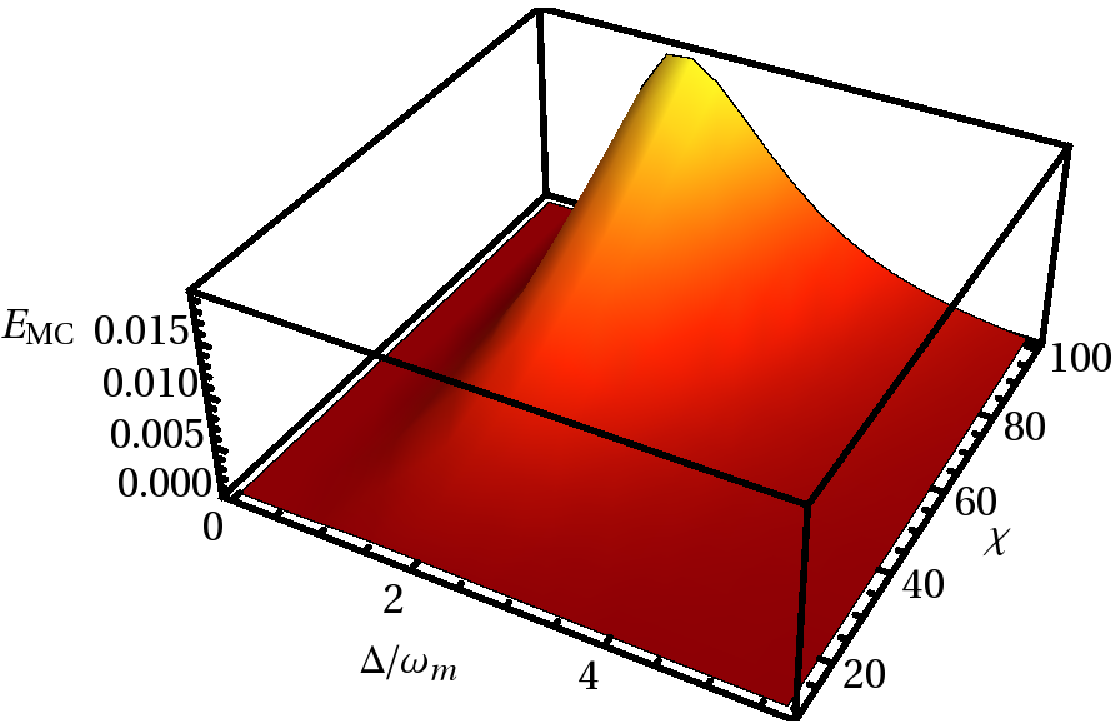}
\includegraphics[scale=0.2]{2c}
\caption{
 {\bf (a)} Logarithmic negativity $E_{AC}$ against $\Delta$ (in units of $\omega_m/2\pi{=}{3}{\times}10^6$s$^{-1}$) and the cavity-mirror coupling rate $\chi$. {\bf (b)} Same as in panel {\bf (a)} but for  $E_{MC}$. We have taken $T{=}10$ $\mu$K, mechanical quality factor $3{\times}10^4$, $m{=}50$ ng, ${\cal R}{=}50$ mW, cavity finesse $F=10^4$ and $\zeta{=}\chi$.  The cavity decay rate is given by $\kappa=\pi c/2 L F$ where $c$ is the speed of light and the cavity length is $L=1$ mm {\bf (c)} Comparison between $E_{AC}$ (solid) and $E_{MC}$ (dotted) against $T$ for $\Delta{=}2\omega_m$ and $\chi{=}100$ s$^{-1}$. Other parameters are as in {\bf (a)}.
}
\label{fig:E5}
\end{figure*}

Any realistic description of the problem at hand should include the most relevant sources of 
noise affecting the overall device, that is energy leakage from the cavity and mechanical Brownian motion at temperature $T$. The resulting open-system dynamics can be efficiently described by means of  a set of Langevin equations for the vector of operators 
$$\hat \phi^T{=}(\hat x,\hat y,\hat q,\hat p,\hat Q,\hat P),$$ 
where 
\begin{equation}
\hat x=\frac{\hat a{+}\hat a^\dagger}{\sqrt 2},
\quad
\hat y=i\frac{\hat a^\dagger{-}\hat a}{\sqrt 2}
\end{equation}
 are the position- and momentum-like quadrature operators of the cavity field.
Under the experimentally reasonable assumption of an intense pumping field, we expand the relevant operators around the respective classical mean values $\phi_{s,j}$  as $\hat \phi_j{=}\phi_{s,j}{+}\delta \hat \phi_j$, where $\delta\hat{\phi}_j$'s are zero-mean fluctuation operators. 
The mean values are easily determined by solving the steady-state Langevin equations, which readily lead to 
\begin{eqnarray}
q_s{=}\frac{\hbar \chi' |\alpha_s|^2}{m\omega_m^2},
\\
Q_s{=}-\frac{ \zeta |\alpha_s|^2}{\Omega},
\end{eqnarray}
with 
$$|\alpha_s|^2{=}\frac{\eta^2}{\Delta^2{+}\kappa^2}$$
 (that also allows us to determine $x_s$ and $y_s$) and the effective cavity detuning 
 \begin{equation}
\Delta{=}\omega_C{-}\omega_L{+}\frac{g^2N_0}{2\Delta_a}{-}\chi' q_s{+}\zeta Q_s.
\end{equation}
  In the rest of the paper, we will use $\Delta$ as a reference parameter, while all the other parameters entering the expression of $\Delta$ remain unspecified.  At this point of our analysis, it is convenient to define dimensionless position and momentum operators for the mirror as $\hat{q}{=}\sqrt{\hbar/m\omega_m}\hat{ \tilde q}$ and  $\hat{p}{=}\sqrt{\hbar m\omega_m} \hat{ \tilde p}$ and cast the Langevin equations for the fluctuations into the form
\begin{equation}
\label{eq:dotphi}
{\partial_t\delta\hat \phi}{=}{\cal K}\delta\hat\phi{+}\hat{\cal N},
\end{equation}
where we have defined the vector of fluctuations 
$$\delta{\hat\phi}^T{=}(\delta\hat x, \delta\hat y, \delta\hat {\tilde q}, \delta\hat {\tilde p}, \delta\hat Q, \delta\hat P),$$
 the noise vector 
$$\hat{\cal N}^T{=}(\!\sqrt{2\kappa}\delta \hat x_{in},\!\sqrt{2\kappa}\delta\hat y_{in}, 0,\frac{\hat\xi}{\sqrt{\hbar m\omega_m}},\!0,\!0)$$
and the coupling matrix
\begin{equation}
K=\begin{pmatrix}
-\kappa & \Delta&0&0&0&0
\\
-\Delta & -\kappa&\sqrt 2\chi' \alpha_s&0&-\sqrt 2\zeta \alpha_s&0
\\
0&0&0&\omega_m&0&0
\\
\hbar \sqrt 2\chi' \alpha_s&0&-\omega_m&-\gamma&0&0
\\
0&0&0&0&0&\Omega
\\
- \sqrt 2 \zeta \alpha_s &0&0&0&-\Omega&0
\end{pmatrix}
\end{equation}
 that is readily determined following Ref.~\cite{PaternostroPRL2010} and depends on the scaled coupling parameter $\chi{=}\chi'\sqrt{\hbar/m\omega_m}$ and the mirror dissipation rate $\gamma$.
The noise vector contains 
the cavity input-noise operators 
\begin{eqnarray}
\delta\hat{x}_{in}&{=}&\frac{\delta\hat{a}_{in}{+}\delta\hat{a}^\dag_{in}}{\sqrt{2}},
\\
\delta\hat{y}_{in}&{=}&i\frac{\delta\hat{a}^\dag_{in}{-}\delta\hat{a}_{in}}{\sqrt{2}},
\end{eqnarray}
such that $\langle\delta\hat a_{in}\rangle{=}0$ and $\langle\delta\hat a_{in}(t)\delta\hat a_{in}^\dagger(t') \rangle{=}\delta(t{-}t')$. Moreover, we have introduced the zero-mean Langevin force operator $\hat\xi$ responsible for Brownian motion of the mechanical system at temperature $T$. 
Small values of $\gamma$ and low operating temperatures allow to retain an effective Markovian description of the mechanical Brownian motion, where~\cite{Giovannetti2001}
\begin{equation}
\langle \xi(t) \xi(t') \rangle{\simeq}\frac{\hbar\gamma m}{\beta_B}\delta(t{-}t')
\end{equation}
with $\beta_B{=}\hbar/(2k_B T)$ and $k_B$ being the Boltzmann constant. Moreover, we have checked the validity of the first-order expansion of the operators (which holds only for small quantum fluctuations) by looking at the stability of the steady-state solution to the Langevin equations, which is guaranteed  when ${\cal K}$ has all negative eigenvalues. 
Using $\Delta$ as a free parameter, the solution is stable when $\Delta{>}0$ and 
$(\chi, \zeta){\ll}\kappa$, which implies that the mirror-BEC-induced back-action is small compared to the cavity dissipation. Such conditions are met across our work.

\section{Stationary entanglement}
\label{sec:ent}
We can now address the entanglement-inference scheme that is the focus of our study. We focus on the $t{\to}\infty$ regime where any transient in the dynamics of the system has faded away and steady-state quantum correlations are set. The main tool of our assessment is the covariance matrix of the state of our hybrid system, which is defined as~\cite{Walls}
\begin{equation}
{\cal V}[i,j]{=}\frac 12 \langle\{ \delta\hat\phi_i, \delta\hat\phi_j\}\rangle{-}\langle \delta\hat\phi_i\rangle \langle \delta\hat\phi_j\rangle
\end{equation}
 and contains full information on a Gaussian state such as ours. Using the Langevin equations one can show that ${\cal V}$ fulfills the equation
\begin{equation}
\label{eq:V}
{\cal K} {\cal V}{+}{\cal V} {\cal K}^T{=-}{\cal D},
\end{equation}
where 
$${\cal D}{=}\text{diag}[\kappa,\kappa,0,\gamma(2\bar n{+}1),0,0]$$ and 
\begin{equation}
\bar n{=}\frac{1}{\exp
(2\beta_B\omega_m
){-}1}
\end{equation}
 is the thermal mean-occupation number of the mechanical state.
This allows us to find ${\cal V}$ for any values of the relevant parameters. We call $A, M, C$ the atomic mode, the mirror one and cavity field respectively and quantify the entanglement between any two modes  $\alpha$ and $\beta$ ($\alpha,\beta{=}A,M,C)$ using the logarithmic negativity  ${\text E}_{\alpha\beta}{=}{-}\log 2\nu$~\cite{logneg}. Here, $\nu$ is the smallest symplectic eigenvalue of the the matrix ${\cal V}'_{\alpha\beta}{=}P{\cal V}_{\alpha\beta}P$, $P{=}\text{diag}(1,1,1,-1)$ performs momentum-inversion in phase-space and ${\cal V}_{\alpha\beta}$ is the reduced covariance matrix of $\alpha$ and $\beta$~\cite{contangle}.

In the following two subsections, we discuss separately the bipartite and the multipartite cases.

\subsection{Bipartite entanglement}

Here we analyze the stationary entanglement in the three possible bipartitions of the system. 
Let us start with the symmetric situation where $\Omega{=}\omega_m$ and where the cavity-mirror coupling equals the cavity-atoms coupling ($\zeta{=}\chi)$. We always restrict ourselves to a stable regime for the linear Langevin system. In this case and for very low initial temperature of the mechanical mode (we take $T{=}10 \mu$K),  we find that entanglement is generated in the stationary state between the Bogoliubov mode and the cavity field as well within the field-mechanical mode subsystem. Due to the symmetric coupling $E_{AC}$ is very similar to $E_{MC}$, as shown in Fig.~\ref{fig:E5}({\bf a-b}). This provides an interesting {\it diagnostic tool}: the inaccessibility of the mirror mode makes the inference of the optomechanical entanglement a hard task needing cleverly designed, although experimentally challenging, indirect methods~\cite{Vitali2007,MauroPRL}. However, in light of the recent demonstration of controllability of intra-cavity atomic systems~\cite{Esslinger2008,Esslinger2007}, we can think of inferring the pure optomechanical entanglement simply by measuring the more accessible correlations between the optical field and the Bogoliubov mode. In the range of parameters considered here, the mirror-atoms entanglement $E_{AM}$ is always zero. This is in contrast with the results in Ref.~\cite{genes}, where the emergence of $E_{AM}$ is due to the use of an effective negative detuning that regulates the free evolution of the collective atomic quadrature. In our case, such evolution is ruled by the frequency $\Omega$, which is always positive. Therefore, the two models can access quite different regions of the parameter-space. In turn, this implies that, while our system does not allow for hybrid atom-mirror entanglement, Genes {\it et al.}in Ref.~\cite{genes} did not achieve the symmetric $E_{MC}{=}E_{AC}$ situation revealed above.

\begin{figure}[t]
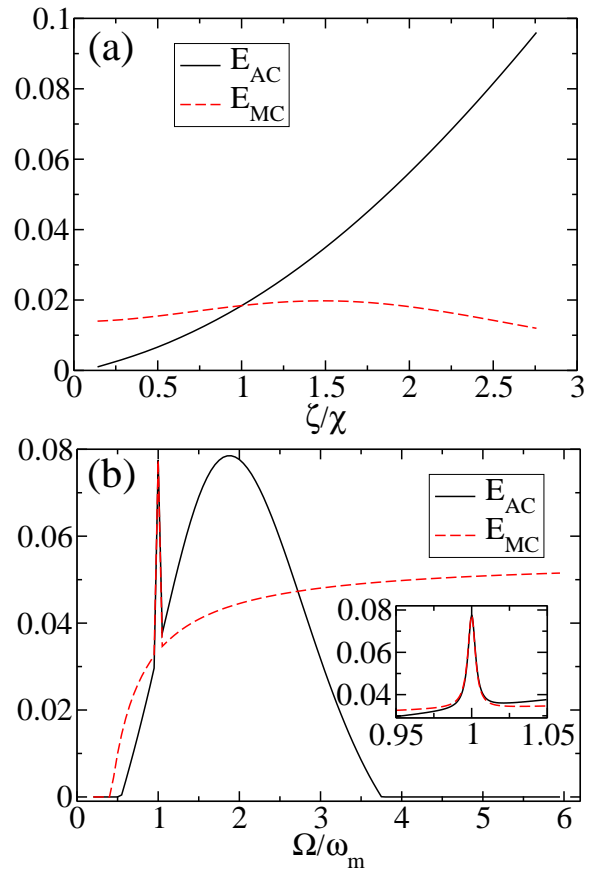

\includegraphics[scale=0.31]{3a}
\includegraphics[scale=0.31]{3b}
\caption{{\bf (a)} Logarithmic negativity $E_{AC}$ (solid) and $E_{MC}$ (dashed) against $\zeta/\chi$ for $\Delta{=}2\omega_m$.
Other parameters are as in Fig.~\ref{fig:E5}. {\bf (b)} We show $E_{AC}$ (line) and $E_{MC}$ (dashed) against $\Omega/\omega_m$. The inset shows the entanglement functions around $\Omega{=}\omega_m$ for
$T{=}1\mu$K and $F{=}4{\times}10^4$. 
}
\label{fig:entasym1}
\end{figure}

We now study how $E_{AC}$ and $E_{MC}$ decay with $T$. The results plotted in Fig.~\ref{fig:E5} {\bf (c)} show that the two entanglements are indistinguishable and therefore the BEC can still be used as an entanglement probe. As expected the entanglement decays when increasing the environment temperature and disappears for $T{>}0.1$mK. Higher critical temperatures for the disappearance of entanglement are found for larger cavity quality factors and more intense pumps.
We consider two different regimes, where we relax the symmetric conditions between the mirror and the Bogoliubov mode. In the first situation, we take $\omega_m{=}\Omega$ and vary the coupling rates. In the second, we take symmetric couplings $\zeta{=}\chi$ and change the frequencies. For  $\omega_m{=}\Omega$, $E_{AC}$ and $E_{MC}$ are strongly affected by the change in $\zeta$ and $\chi$. As shown in Fig.~\ref{fig:entasym1} {\bf (a)}, $E_{AC}$ grows continuously for larger $\zeta$ while $E_{MC}$ decreases slightly.
In this regime, it is clear that small inaccuracies of the ratio $\zeta/\chi$ do not affect the mirror-cavity entanglement, therefore confirming the role of the BEC as a minimal disturbance probe.
A more involved situation occurs when we keep $\omega_m$ fixed and change $\Omega$. We take $T{=}1\mu$K, cavity finesse $F{=}4{\times}10^4$ and find a sharp peak at $\Omega{=}\omega_m$ where $E_{MC}{=}E_{AC}$, see [Fig.~\ref{fig:entasym1} {\bf (b)}]. While $E_{MC}$ increases slowly with $\Omega/\omega_m$ (the Bogoliubov mode goes out of resonance from the cavity and decouples from the rest of the system), $E_{AC}$ reaches its  maximum at $\Omega/\omega_m{\approx}2$.
These results show that by changing the Bogoliubov frequency, for example varying the longitudinal trapping frequency, the BEC acts as a switch for the mirror-cavity entanglement, inhibiting or enhancing it.

\subsection{Genuine multipartite entanglement}
\label{sec:multi}

\begin{figure}[t]
\includegraphics[scale=0.35]{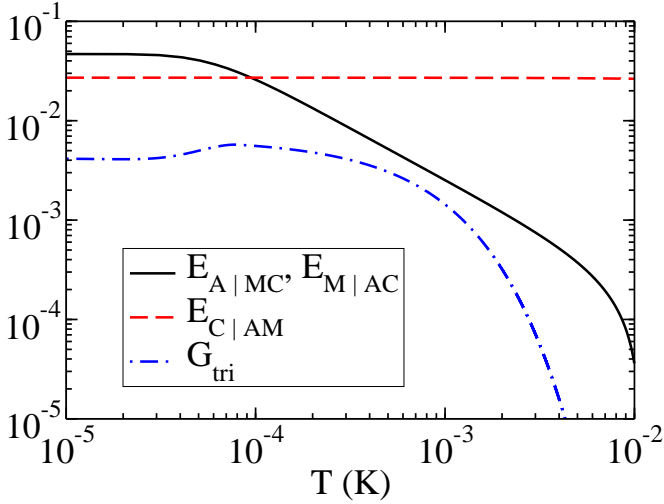}
\caption{We show $E_{i|jk}$  ($i,j,k{=}A,M,C$) and the genuine tripartite entanglement $G_{tri}$ against $T$ for $\Omega{=}\omega_m$ and $\chi{=}\zeta$. We have $E_{A|MC}{=}E_{M|AC}$ (solid), $E_{C|AM}$ (dashed) and $G_{tri}$ (dot-dashed). 
Other parameters are as in Fig.~\ref{fig:E5}.
}
\label{fig:tripartite}
\end{figure}

We now study the existence of genuine multipartite entanglement in our system looking at the logarithmic negativity in each one-vs-two-mode bipartition. According to Ref.~\cite{giedke}, if in a tripartite state all such bipartitions are inseparable, genuine multipartite entanglement is shared. For the systems at hand, we indeed find inseparability of the bipartitions $A|MC$, $M|AC$ and $C|AM$ [see Fig.~\ref{fig:tripartite}], which strongly confirms the presence of tripartite entanglement up to very high temperatures. Evidently, multipartite entanglement persists up to  $T{\sim}{0.01}$K, which is much larger than the critical one for the disappearance of bipartite entanglement (${\sim}8\times10^{-5} K$). Such a result spurs the quantitative study of the genuine tripartite-entanglement content of the overall state~\cite{contangle}. For pure multipartite states, this is based on {monogamy inequalities} valid for the {squared logarithmic negativity}, which turns out to be a proper entanglement monotone. For mixed states, the convex-roof extension of such a measure is required, albeit restricted to the class of Gaussian states. More explicitly, we aim at determining
\begin{equation}
\label{cota}
{\text G}_{\text{tri}}{=}\min_{\Pi({ijk})}[{{\text G}_{i|jk}{-}{\text G}_{ij}{-}{\text G}_{ik}}]
\end{equation}
with $\Pi(ijk)$ the permutation of indices $i,j,k{=}A,M,C$ and $\text{G}$ the convex roof of the squared logarithmic negativity for a given bipartition of the system. In general, the evaluation of $\text{G}_{\text{tri}}$ is very demanding. However, for $\omega_m{=}\Omega$ and $\chi{=}\zeta$, the covariance matrix $\mathcal V$ is symmetric under the permutation of $A$ and $M$, which greatly simplifies the calculations: The residual tripartite entanglement can be thus determined against the effects of $T$, showing a non monotonic behavior. To understand this, we take $i{=}C$, $j{=}M$ and $k{=}A$ (any other combination could be considered). Similar to $E_{AC}$ and $E_{MC}$, $\text{G}_{C|A,M}$ decays very quickly as $T$ increases, thus biasing the competition between ${\text G}_{C|MA}$ and ${\text G}_{C|A}+{\text G}_{C|M}$. Analogously to $E_{C|AM}$, ${\text G}_{C|A}+{\text G}_{C|M}$ decreases very slowly with $T$, thus determining an overall increase of the residual entanglement. However, as $T$ is raised further, the system tends toward two-mode biseparability and any tripartite entanglement is washed out [see Fig.~\ref{fig:tripartite}].

\section{Probing experimentally the stationary entanglement}
\label{sec:det}
We can now discuss the direct observation of the entanglement between the Bogoliubov mode and the cavity field. The idea is to shine the BEC with a probe laser in a standing wave
~\cite{RitschNatPhys}  forming a small angle with respect to the cavity axis, as sketched in Fig.~\ref{fig:sketch}.
For an off-resonant probe field with polarization perpendicular to that of the primary cavity field, 
the additional Hamiltonian terms read
\begin{equation}
\hat{\cal H}_P{=}\hbar\left[\Delta_P{+} U_P \int d^3 {\bf r} \cos^2(k_c x{+}k_z z)
\hat \Psi^\dagger\hat \Psi\right]\hat a_P^\dagger \hat a_P,
\end{equation}
where $\Delta_P$ is the detuning of the probe field from the pumping laser,
$\hat a_P$ is the corresponding annihilation operator, $U_P{=}g_P^2/\Delta_{AP}$ is the light-atom coupling constant, $k_z$ is the wave-vector of the probing field along a direction orthogonal to the cavity axis and $\hat\Psi$ is the condensate field operator. For a BEC  strongly confined in the plane orthogonal to the cavity axis, we can neglect excitations of transverse modes and factorize the field operator as 
\begin{equation}
\label{eq:psifactor}
\Psi{=}\psi(x) f(y) f(z),
\end{equation}
where $f(y)$ is a function peaked around the cavity axis and characterized by a width $\sigma$. We then expand $\psi(x)$ as~\cite{StringariPitaevskii}
\begin{equation}
\psi(x){=}\sqrt{N_0} \psi_0(x){+}\sum_{k>0,\sigma=\pm}[u_{k\sigma}(x) c_{k\sigma}{-}v_{k\sigma}^*(x) c_{k\sigma}^\dagger].
\end{equation}
We assume that the width of the BEC in the transverse direction is much smaller than the periodicity of the probe field along the $z$ direction. This implies that transverse excitations are suppressed.
Neglecting terms ${\cal O}(k_z^2\sigma^2)$ we get
\begin{equation}
\hat{\cal H}_{AP}{\simeq}\hbar\left(\Delta_P{+}\frac{N_0 U_P}{2}{+}\zeta_P \hat Q \right)\hat a_P^\dagger \hat a_P{-}i\hbar \eta_P(\hat a_P{-}\hat a^\dagger_P),
\end{equation}
where $\zeta_P$ is the same as $\zeta$ with $U_0{\rightarrow}U_P$.
The Langevin equation for the probe field reads
\begin{equation}
{\partial}_t \hat{a}_P{=}\eta_P{-}i\left[\Delta_P{+}\frac{N_0 U_P}{2}+\zeta_P \hat Q\right] \hat a_P
{+}\sqrt{2\kappa} \hat a_{Pin}{-}\kappa \hat a_P.
\end{equation}
Assuming again an intense pump field, the steady intensity of the probe field is
\begin{equation}
|\alpha_P|^2{=}\frac{\eta_P^2}{\tilde\Delta_P^2+\kappa_P^2},
\end{equation}
where 
$$\tilde\Delta_P{=}\Delta_P{+}\frac{U_P N_0}{2}{+}\zeta_P Q_s.$$
We also assume that $\zeta_P{\ll}\Omega,\zeta$ and $\alpha_P{\ll}\alpha_S$.
The equation of motion for the fluctuations of the probe field is
\begin{equation}
\partial_t\delta \hat a_P{=}-i \tilde \Delta_P \delta \hat a_P{-}i\zeta_P\alpha_p\delta \hat Q{+}\sqrt{2\kappa}\delta \hat a_{Pin}{-}\kappa_P\delta \hat a_P.
\end{equation}
In order to map the Bogoliuobov mode onto the probe field, and following the same technique as in~\cite{Vitali2007}, we choose $\tilde\Delta_P{=}\Omega{\gg}\kappa_P, \zeta_P\alpha_P$. The equations of motion for the slowly varying variables $\hat{\tilde o}(t){=}e^{i\Omega t} \hat o(t)$ thus read
\begin{equation}
\partial_t\delta\hat{\tilde a}{=}-i\frac{\zeta_P \alpha_P}{\sqrt 2}\hat{ \tilde c}{+}\sqrt{2\kappa}\delta\hat{\tilde a}_{Pin}{-}\kappa_P\delta \hat{\tilde a}_P.
\end{equation}
If the decay of the probe cavity is faster than the dynamics of the Bogoliubov modes, the probe field follows adiabatically the dynamics of the latter. Using the cavity input-output relations~\cite{Walls} we get
\begin{equation}
\delta\hat{\tilde a}_{Pout}{=}{-}i\frac{\zeta_P \alpha_P}{\sqrt \kappa_P}\hat{\tilde c}{+}\delta\hat{\tilde a}_{Pin}
\end{equation}
which shows how the Bogoliubov mode is mapped onto the output probe field. To measure the entanglement between the field of the primary cavity and the Bogoliubov mode one can homodyne the cavity field and rotate the quadrature of the probe. Analogously, one could map $\hat{q}$ onto a further field so that the light-matter correlations are changed into amenable light-light ones~\cite{Vitali2007}.

\section{Conclusions}
\label{sec:conclusion}
{We have demonstrated the proof of principle of the use of a BEC as a diagnostic tool to determine the elusive mirror-light entanglement in a hybrid optomechanical device.
}
No extra assumption is required for such an inference protocol that can be demonstrated using current experimental abilities~\cite{Esslinger2008,Stamper2008}. The entanglement structure within the system is intriguing and includes the case of tripartite inseparability, resilient to temperatures of up to a few milli-Kelvin and fully experimentally characterizable using existing technology. This demonstrates the viability of our scheme for diagnostics which contributes to the study of a complex three-body interaction exhibiting genuine quantum features at the macroscale.

{\it Acknowledgments.--} We thank P. Massignan, A. Sanpera and J. Stasinska for discussions. GDC is supported by the Spanish MICINN (Juan de la Cierva, FIS2008-01236 and QOIT-Consolider Ingenio 2010), Generalitat de Catalunya Grant No. SGR2009-00347. GMP acknowledges EUROTECH. MP thanks EPSRC (EP/G004579/1) for financial support.



\end{document}